\documentclass[a4paper]{jpconf}
\usepackage{graphicx}
\usepackage{subfigure,graphicx,slashed,color}
\usepackage{wrapfig}
\usepackage{xspace,colortbl,times}
\usepackage{lmodern}
\usepackage{textcomp}
\usepackage[english]{babel}
\usepackage{blindtext}
\usepackage{tabularx}
\usepackage{amsmath}
\usepackage{amssymb}
\usepackage{fancyvrb}
\usepackage{fancyhdr}
\usepackage{bm}
\usepackage{framed}
\usepackage{booktabs}
\usepackage{alltt}
\usepackage{listings}
\usepackage{lscape}
\usepackage{rotating}
\usepackage{booktabs}
\usepackage{array}
\usepackage{rotating}

\usepackage[left]{lineno}
\usepackage{blindtext}

\begin{document}


\newcommand{\pip}{\ensuremath{\pi^{+}}}
\newcommand{\pim}{\ensuremath{\pi^{-}}}
\newcommand{\piz}{\ensuremath{\pi^{0}}}
\newcommand{\kap}{\ensuremath{{\rm K}^{+}}}
\newcommand{\kam}{\ensuremath{{\rm K}^{-}}}
\newcommand{\pbar}{\ensuremath{\rm\overline{p}}}
\newcommand{\sqrtS}{\ensuremath{\sqrt{s}}}
\newcommand{\pp}{\ensuremath{\mathrm {p\kern-0.05emp}}}
\newcommand{\PbPb}{\ensuremath{\mbox{Pb--Pb}}}
\newcommand{\pPb}{\ensuremath{\mbox{p--Pb}}}
\newcommand{\dEdx}{\ensuremath{\mathrm{d}E/\mathrm{d}x}}
\newcommand{\dndy}{\ensuremath{\mathrm{d}N/\mathrm{d}y}}

\title{Measurement of open heavy-flavour production as a function of charged-particle multiplicity with ALICE at the LHC}

\author{Sibaliso Mhlanga\\ University of Cape Town, Rondebosch, Cape Town, South Africa \\ iThemba LABS, Sommerset West, Western Cape }

\address{For the ALICE Collaboration}

\ead{sibahso.mhlanga@cern.ch}

\begin{abstract}
Heavy quarks are produced in the early stages of ultra-relativistic hadron collisions via hard scatterings and are an important tool for studying different aspects of Quantum Chromodynamics (QCD) in hadronic collisions. Charged-particle multiplicity gives information on the global characteristics of the event and could be used to characterize particle production mechanisms. In hadronic collisions at Large Hadron Collider (LHC) energies, there is a significant contribution of multi-parton interactions. The measurement of heavy-flavour yields as a function of charged-particle multiplicity gives insight into the mechanisms influencing their production in hadronic collisions at these energies and it is a tool to test the possible influence of multi-parton interactions. Furthermore, the charged-particle multiplicity dependence of open heavy flavours is used to test the ability of QCD theoretical models to describe the data. In ALICE, heavy-flavour production is measured via the hadronic and semi-leptonic decay channels (electrons at central rapidity and muons at forward rapidity). Charged-particle multiplicity is measured at central and forward rapidity. We will present the results on open heavy-flavour production as a function of the charged-particle multiplicity in \pp\ and \pPb\ collisions. Results will be compared to quarkonia measurements as well as theoretical model calculations.
\end{abstract}

\section{Introduction}
Heavy quarks (charm and beauty) are abundantly produced at the LHC \cite{LHC} in the early stages of hadronic collisions via hard scattering processes and  experience the full evolution of the collision. The measurement of heavy flavours in \pp\ collisions can be used to test pQCD calculations. In addition, they provide an  essential baseline for the studies in nuclear (\pPb\ and \PbPb) collisions. Their production in nuclear collisions is modified by cold nuclear matter effects (CNM) such as shadowing and energy loss \cite{eloss, cnm}. The knowledge of these effects is fundamental for understanding the interactions of heavy quarks with the deconfined medium formed in heavy-ion collisions where the modified transverse momentum distribution can be used to infer the medium's characteristics. Heavy quark production is experimentally accessible through the measurement of heavy-flavour hadrons (D or B hadrons) via their decay products. \newline
 
 The charged-particle multiplicity is defined as the average number of charged particles produced per event. It is a key observable for multi-parton interactions (MPI) in high-energy hadronic collisions \cite{mpi1, mpi2}. The charged-particle multiplicity may be used to test models of particle production such as PYTHIA \cite{pythia} and PHOJET \cite{phojet}. Finally, the charged-particle multiplicity dependence of open heavy-flavour yields provides insight into mechanisms of particle production in hadronic collisions. In this report we present the relative yield of D mesons as a function of the relative charged-particle multiplicity in \pp\ collisions at $\sqrt{s}$ = 7 TeV  \cite{ppcollisions} and in \pPb\ collisions at $\sqrt{s_{NN}}$ = 5.02 TeV \cite{pPbcollisions}. \newline
 
At the LHC, ALICE \cite{alice} comprises central detectors, used for the measurement of charged particles as well as photons and a forward muon spectrometer which measures muons. The L3 solenoid magnet provides a field of 0.5 T to the central barrel detectors covering the rapidity -0.9 $<$ $\textit{y}$ $<$ 0.9. The Silicon Pixel Detector (SPD) is used for the determination of the interaction vertex as well as the measurement of charged-particle multiplicity. Particle identification information is given by the Time Of Flight (TOF) and the Time Projection Chamber (TPC).   Two arrays of scintillator detectors (V0A and V0C)  placed on both sides of the interaction point  provide centrality in \PbPb\ collisions as well as trigger information. They are also used for multiplicity determination. The forward muon spectrometer covers the rapidity range -4 $<$  $\textit{y}$ $<$ -2.5. It consists of a composite absorber, a dipole magnet, five tracking stations, a muon filter and two trigger stations. The absorber reduces background muons from the decays of pions and kaons. The dipole magnet  provides a horizontal magnetic field perpendicular to the beam axis and is used for charge and momentum determination. The tracking stations are used to determine the trajectories of the muons traversing the detector. The muon filter is mounted at the end of the spectrometer and filters all particles with momentum $\textit{p}$ $<$ 4 GeV/c. The trigger stations are used for muon identification and triggering.\newline
 

The reconstruction of D mesons is done via their hadronic decay channels: D$\textsuperscript{0}$ $\rightarrow$ \kam\pim (branching ratio, BR = 3.88$\pm$0.05 \%), D$\textsuperscript{+}$ $\rightarrow$ \kam\pip\pip (BR = 9.13$\pm$0.19 \%), D$\textsuperscript{*+}$ $\rightarrow$ D$\textsuperscript{0}$\pip\ (BR = 67.7$\pm$0.05 \%) followed by D$\textsuperscript{0}$ $\rightarrow$ \kam\pip \cite{branchingratios} and their charge conjugates at central rapidity. Further details on D meson and charged-particle multiplicity measurements are reported in \cite{ppcollisions} and \cite{ppcollisions1}. The measurement of prompt and non-prompt (produced via the decay of beauty hadrons) J/$\psi$ $\rightarrow$ e$\textsuperscript{+}$e$\textsuperscript{+}$ (BR = 5.97$\pm$0.03 \%) and J/$\psi$ $\rightarrow$ $\mu$$\textsuperscript{+}$$\mu$$\textsuperscript{+}$ (BR = 5.96$\pm$0.03 \%) at central and forward rapidity are reported in \cite{jpsi}.
 \
 \section{Results}
\subsection{$\textbf{D-meson production in $\textit{pp}$ collisions}$}
\begin{figure}[!h]
		\centering
		
			\subfigure[]
			{
				\includegraphics[width=200px,height=180px]{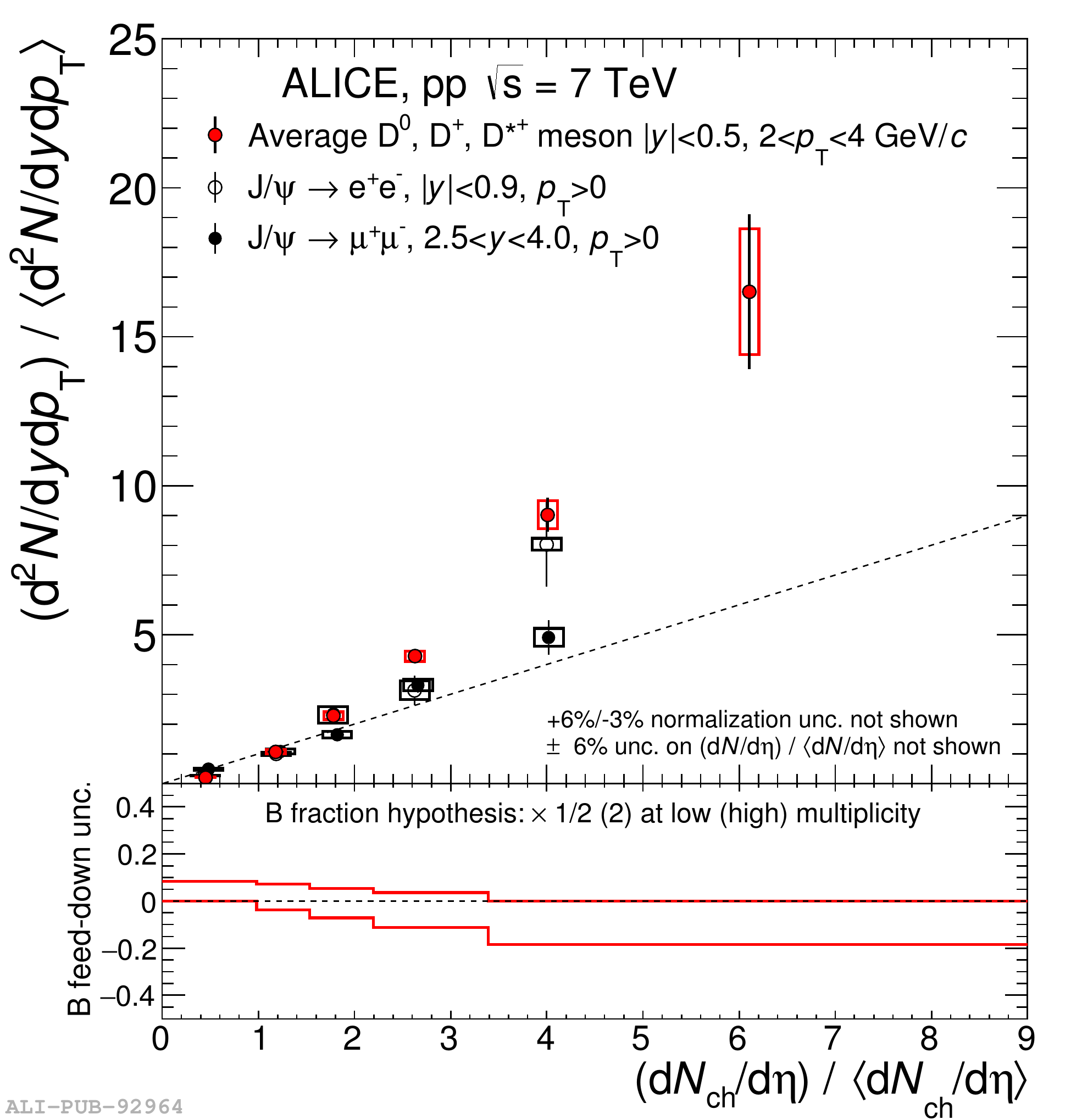}
				\label{fig:first_1}
			}
			\subfigure[]
			{
				\includegraphics[width=200px,height=180px]{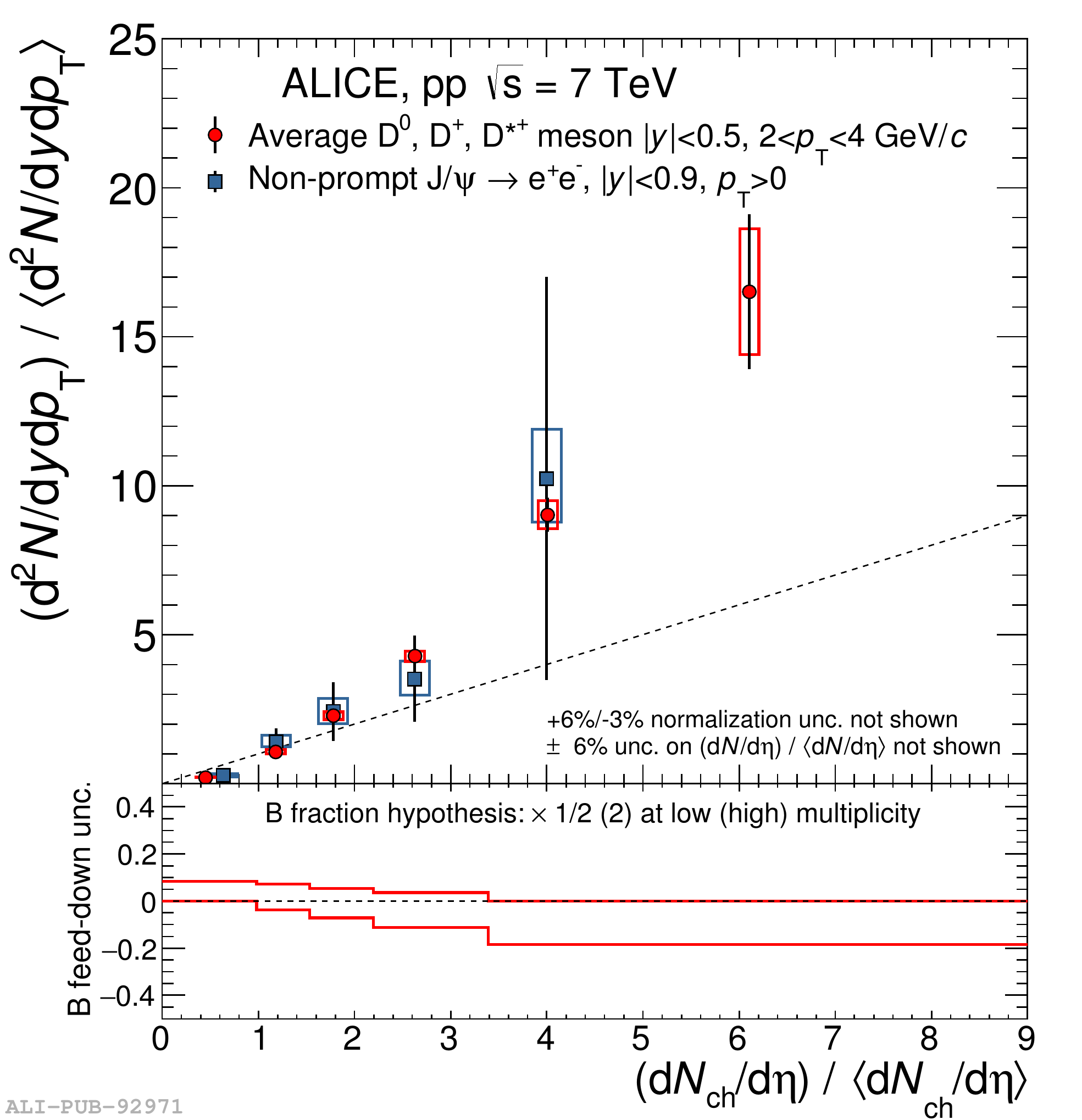}
				\label{fig:first_2}
			}
			\caption{Average D meson yield as a function of the relative charged-particle multiplicity in \pp\ collisions at $\sqrt{s}$ = 7 TeV, measured at central rapidity in the interval 2 $<$ $\textit{p}$$_\text{T}$ $<$ 4 GeV/c compared to the yields of inclusive J/$\psi$ (a) and non-prompt J/$\psi$ (b) for $\textit{p}$$_\text{T}$ $>$ 0 GeV/c \cite{ppcollisions, jpsi}. The vertical bars and the boxes represent the statistical and systematic uncertainties respectively. The bottom panels show the B feed-down fraction uncertainty.}
			\label{results1}
		
	\end{figure}

The results of the relative yields (ratio between the yield in a given multiplicity interval normalised to the
multiplicity-integrated one) of D mesons, D$\textsuperscript{0}$, D$\textsuperscript{+}$, D$\textsuperscript{*+}$, measured at central rapidity as a function of relative charged-particle multiplicity (multiplicity of charged particles normalised to the average value) in \pp\ collisions at $\sqrt{s}$ = 7 TeV are presented in Figure \ref{results1}. In Figure \ref{fig:first_1}, the prompt D-meson relative yield is compared with that of inclusive J/$\psi$ as a function of the relative multiplicity. The yields of D mesons and inclusive J/$\psi$ show a similar increase with the multiplicity, in particular for J/$\psi$ measured at central rapidity. In Figure \ref{fig:first_2} the yield of D mesons is compared to that of non-prompt J/$\psi$ (produced in the decay of beauty hadrons), which provides an indirect measurement of the multiplicity dependence of beauty hadrons. The trend is similar to that observed in Figure \ref{fig:first_1}. We observe no difference, within uncertainty, in the multiplicity dependence of charm and beauty hadron yields at central rapidity.



		\begin{figure}[!h]
			\centering
			
\end{figure}

\subsection{$\textbf{D-meson production in \pPb\ collisions}$}
\begin{figure}[!h]
	\centering
				\includegraphics[width=200px,height=180px]{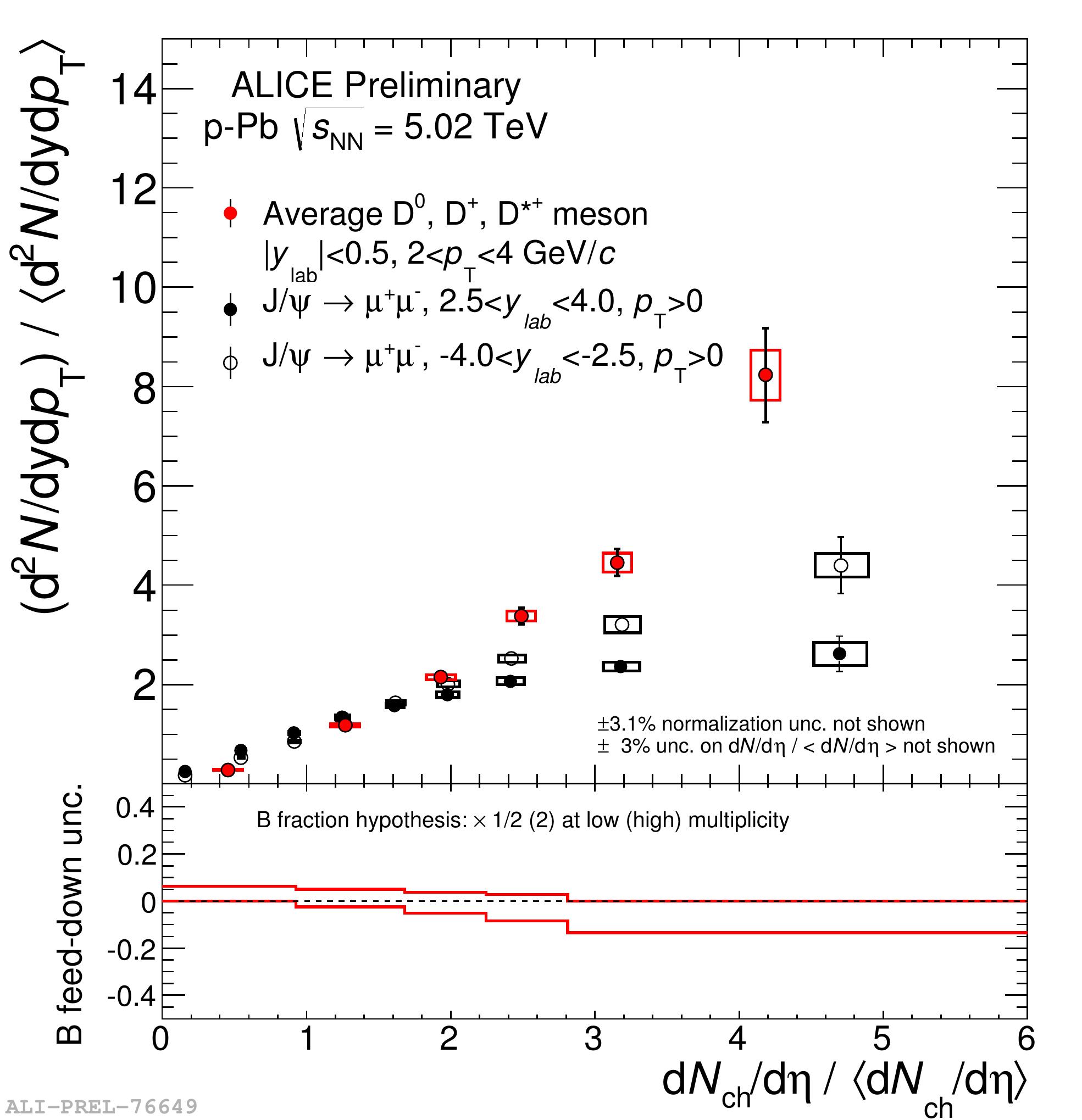}
				\label{fig:first_4_1}
		\caption{Relative yields of the average of the D-meson species and J/$\psi$ in the forward (p going) and backward (Pb going) rapidity as a function of the relative charged-particle multiplicity \cite{pPbcollisions, jpsi}. The vertical bars and the boxes represent the statistical and systematic uncertainties respectively. The bottom panel shows the B feed-down fraction uncertainty.}
		\label{results3}
\end{figure}

Figure \ref{results3} shows the prompt D-meson yield as a function of the relative multiplicity measured at central rapidity (-0.9 $<$ $\textit{y}$ $<$ 0.9) compared with J/$\psi$ in \pPb\ collisions at $\sqrt{s_{NN}}$ = 5.02 TeV at forward (p going) and backward (Pb going) rapidity. The yield of D mesons increases faster than the J/$\psi$ yield towards high multiplicity, particularly for J/$\psi$ measured at forward rapidity.
\subsection{$\textbf{Comparison of D-meson measurements in $\textit{pp}$ and \pPb\ collisions}$}
\begin{figure}[!h]
	\begin{center}
		\includegraphics[width=200px,height=180px]{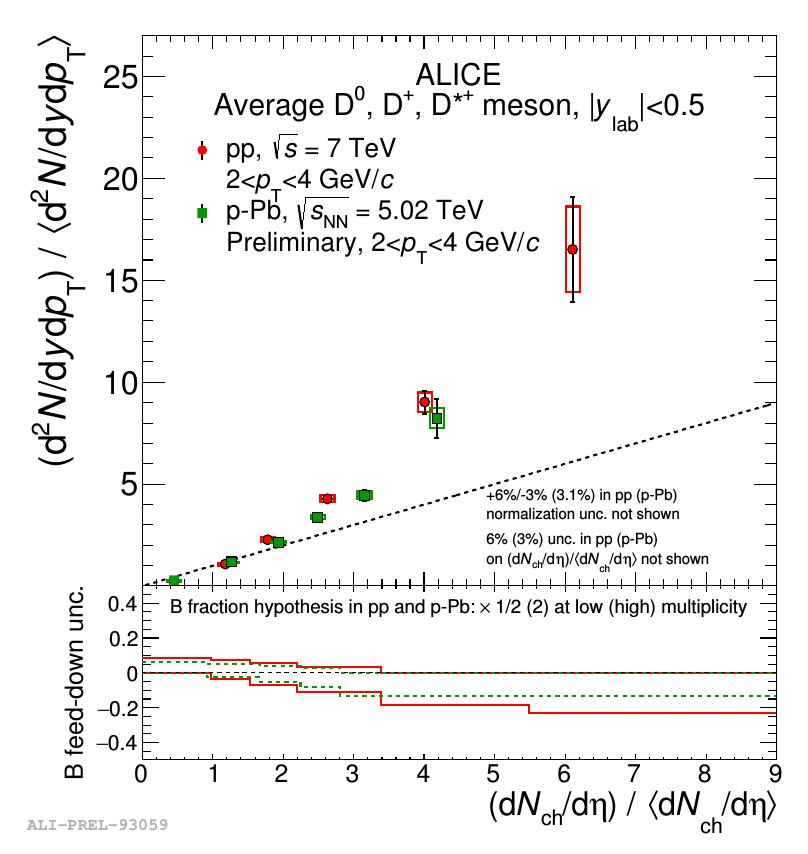}
				\caption{The relative yields of the averages of the three D-meson species as a function of the relative charged-particle multiplicity in \pp\  and \pPb\ collisions at  $\sqrt{s}$ = 7 TeV and $\sqrt{s_{NN}}$ = 5.02 TeV, respectively \cite{pPbcollisions}. The vertical bars and the boxes represent the statistical and systematic uncertainties respectively. The bottom panel shows the B feed-down fraction uncertainty.}
		\label{results5}
	\end{center}
\end{figure}

The yield of D mesons in \pp\ collisions at  $\sqrt{s}$ = 7 TeV is compared to the yield in \pPb\ collisions at $\sqrt{s_{NN}}$ = 5.02 TeV as a function of charged-particle multiplicity in Figure \ref{results5}. The D-meson yield increases faster than linear with multiplicity for both \pp\ and \pPb\ collisions. The increase is independent of the colliding system. 
\subsection{$\textbf{Comparison of results with theoretical models}$}
\begin{figure}[!h]
	\centering
	\subfigure[]
	{
		\includegraphics[width=190px,height=190px]{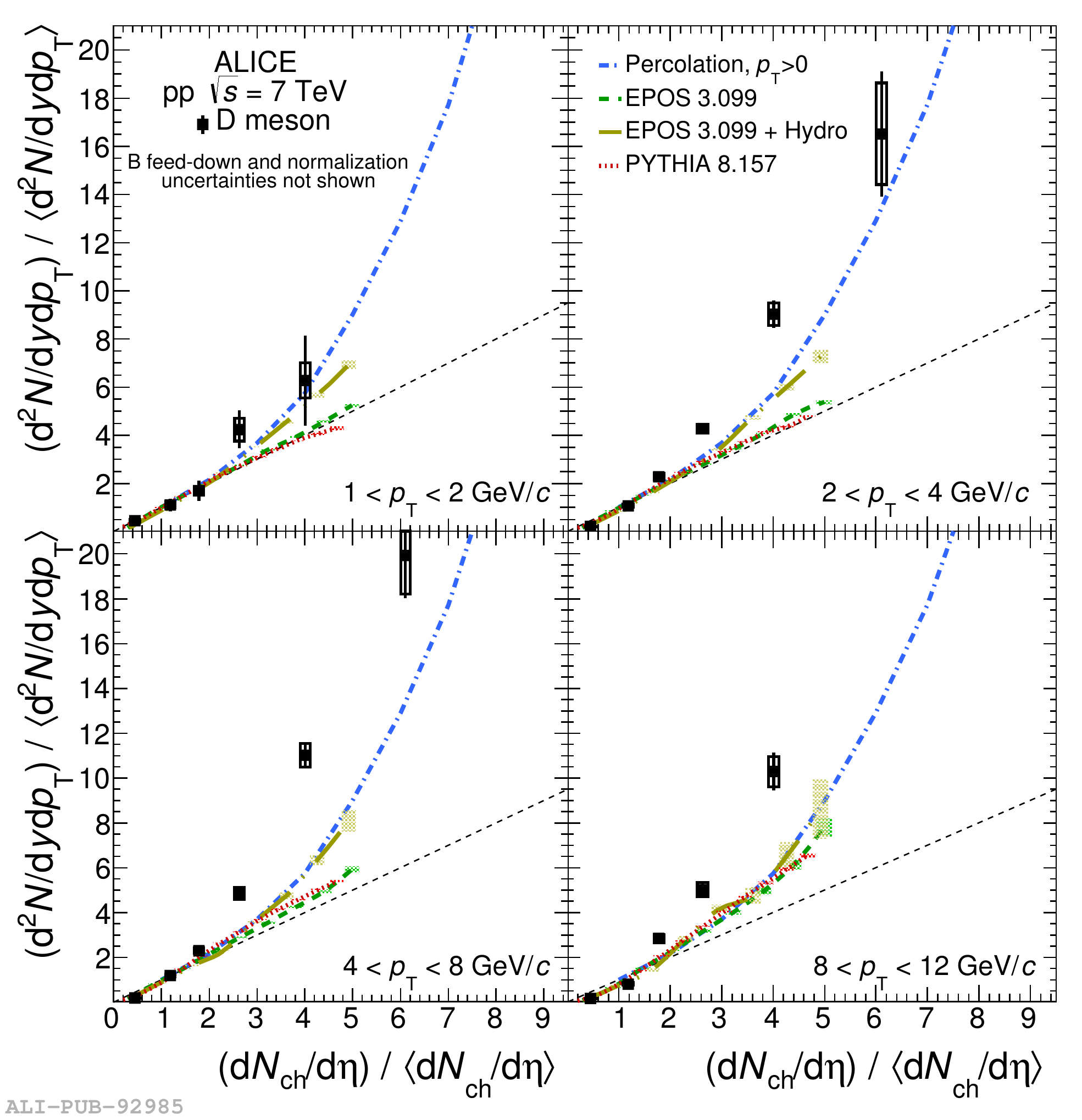}
			\label{fig:first_5}
		}
			\subfigure[]
			{
		\includegraphics[width=190px,height=190px]{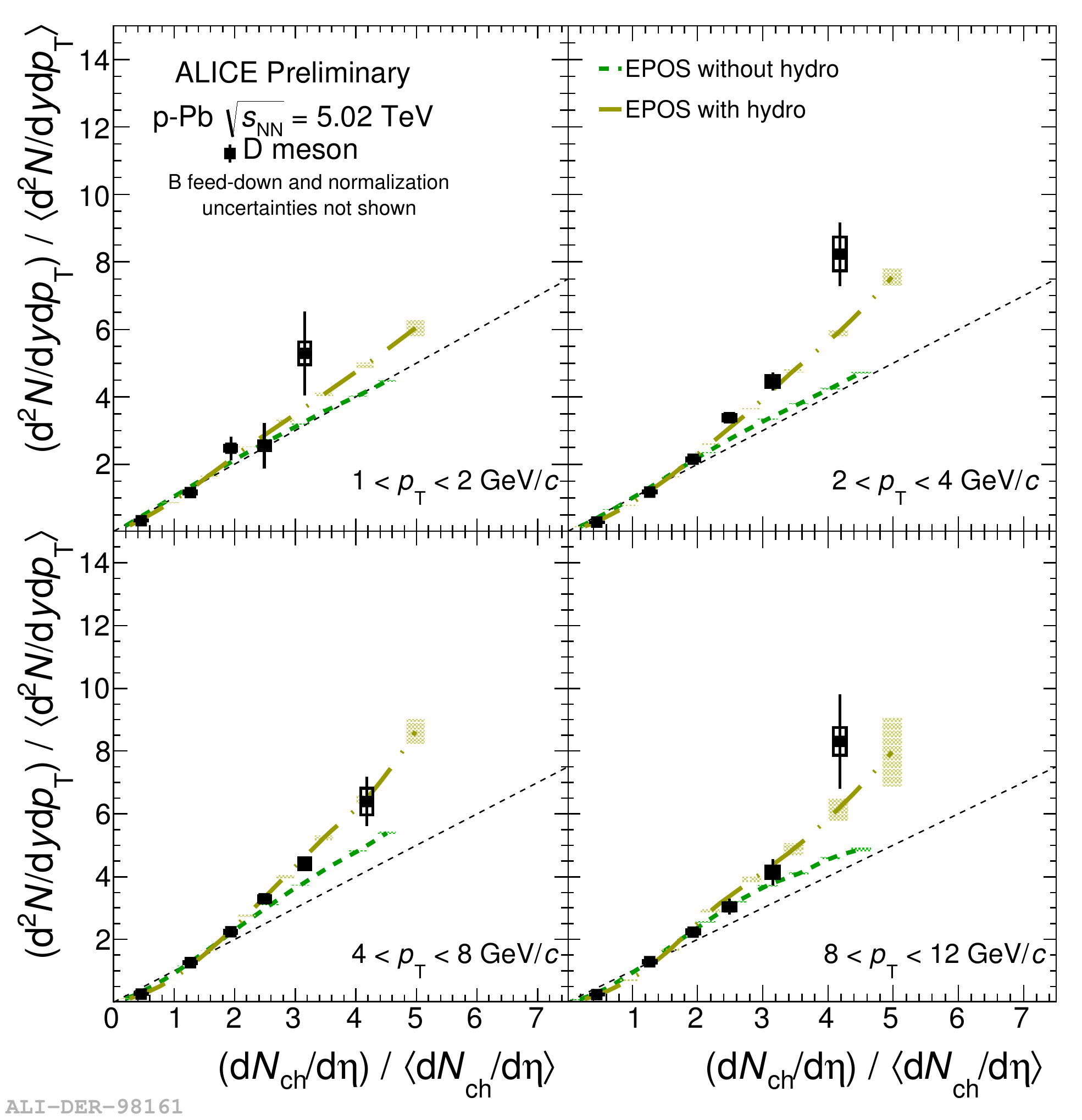}
		
			\label{fig:first_6}
		}
		\caption{Relative yields of D mesons in different $\textit{p}$$_\text{T}$ intervals as a function of the relative charged-particle multiplicity in \pp\ collisions at $\sqrt{s}$= 7 TeV (a) and \pPb\ collisions at $\sqrt{s}$ = 5.02 TeV (b). The lines represent various theoretical models \cite{percolation} - \cite{pythia8}.}
		\label{results6}
	\end{figure}

The comparison of the measurements in \pp\ and \pPb\ collisions with theoretical calculations is shown in Figure \ref{results6}. In \pp\ collisions the yield of D mesons as a function of the charged-particle multiplicity is compared with theoretical predictions based on percolation \cite{percolation, percolation1}, EPOS \cite{epos} with and without hydrodynamical effects and PYTHIA 8 \cite{pythia8}.  All the models predict an increase of the relative yield of D mesons with the relative multiplicity for both \pp\ and \pPb\ collisions. However, percolation and EPOS+hydro predict a faster than linear increase while PYTHIA 8 and EPOS are consistent with the data at lower multiplicity. In \pPb\ collisions, the results are compared with EPOS+hydro and EPOS. Both models predict an increase of the relative yield with the multiplicity. EPOS+hydro predicts a faster than linear increase.



\section{Conclusion}
The ALICE Collaboration measured the relative yield of D mesons as a function of the charged-particle multiplicity in \pp\ and \pPb\ collisions and compared the results with analogous measurements of inclusive and non-prompt  J/$\psi$. In both colliding systems the relative yield of D mesons increases with multiplicity and the increase is faster than linear. The increase is independent of the $\textit{p}$$_\text{T}$ interval considered. J/$\psi$ yields show a similar increase in \pp\ collisions, however in  \pPb\ collisions D-meson yields increase faster than J/$\psi$, particularly at forward rapidity. The results were also compared with theoretical calculations. All models considered in the study predict an increase of heavy-flavour yields with multiplicity. The multiplicity dependence of D-meson production is described better by calculations that include hydrodynamics, especially at higher multiplicity.

\end{document}